\documentclass[reqno,prl,showpacs,twocolumn,superscriptaddress]{revtex4-1}
\usepackage{graphicx}
\usepackage{color}
\usepackage{amsmath}
\usepackage{amssymb}
\usepackage{changes}
\usepackage[cal=boondoxo]{mathalfa}

\newcommand{\cN}{{\cal N}}

\newcommand{\la}{{\mathcal{A}}}

\newcommand{\id}{\textrm{d}}
\newcommand{\ve}{\varepsilon}

\def\bea{\begin{eqnarray}}
\def\eea{\end{eqnarray}}
\def\ba{\begin{array}}
\def\ea{\end{array}}

\def\la{\langle}
\def\ra{\rangle}

\begin{document}

\title{Extrapolation to nonequilibrium from coarse grained response theory}

\author{Urna Basu}
\affiliation{SISSA - International School for Advanced Studies and INFN,  Trieste, Italy}
\affiliation{LPTMS, CNRS, Univ. Paris-Sud, Universit\'{e} Paris-Saclay, 91405 Orsay, France}

\author{Laurent Helden}
\affiliation{2. Physikalisches Institut, Universit\"at  Stuttgart, 70550 Stuttgart, Germany}

\author{Matthias Kr\"uger}
\affiliation{4th Institute for Theoretical Physics, Universit\"at
Stuttgart, Germany} 
\affiliation{Max Planck Institute for Intelligent Systems, 70569 Stuttgart, Germany}
\affiliation{Institute for Theoretical Physics, Georg-August-Universit\"{a}t G\"{o}ttingen, 37077 G\"{o}ttingen, Germany}        

\begin{abstract}
Nonlinear response theory, in contrast to linear cases, involves (dynamical) details, and this makes application to many body systems challenging. From the microscopic starting point we obtain an exact response theory for a small number of coarse grained degrees of freedom. With it, an extrapolation scheme uses near-equilibrium measurements to predict far from equilibrium properties (here, second order responses). Because it does not involve system details, this approach can be applied to many body systems. It is illustrated in a four state model and in the near critical Ising model.
\end{abstract}

\maketitle

Understanding properties of nonequilibrium systems is an ambitious goal of modern statistical physics \cite{Seifert12}, and here, the fluctuation dissipation theorem (FDT) is of fundamental importance: It relates the linear response of a system to its thermal fluctuations in the equilibrium state \cite{Kubo54, Kubo12}. This insight is of practical benefit in solid state physics \cite{Phillips} as well as in classical systems.

The FDT holds close to equilibrium, and extending it to far from equilibrium has been the subject of intense research. The case of small perturbations of far-from-equilibrium states has been analyzed in various works \cite{Seifert12,Ruelle98,Harada05, Speck06, Blickle07, Chetrite08, MARCONI2008111, Baiesi09, Prost09, Krueger09,PhysRevLett.112.140602}.
Another direction aims at finding the nonlinear response, i.e., the response to strong perturbations. The derived formulas relate response functions to nonequilibrium correlation functions \cite{Yamada67,Evans88,Fuchs05}, or to (higher order) correlation functions evaluated in equilibrium \cite{Kubo54,Semerjian04,PhysRevB.72.064204,  Andrieux07,Lippiello08,Colangeli11,Lucarini12,Basu2015}. The latter concept has been applied experimentally only recently \cite{Helden16}, where the second order response was obtained from an equilibrium measurement.

Extensions of FDT to far from equilibrium cases are typically plagued by a property, which is deeply inherent to nonequilibrium physics: Their application requires information about the interactions and dynamics of the system, so that in principle all degrees of freedom (or their nonequilibrium-distributions) have to be tracked during the measurement (see discussions in Refs.~\cite{Basu2015, Seifert12}). This statement may be exemplified for colloidal particles, investigated in Ref.~\cite{Helden16}: To apply second order response theory, the interaction potential of the particles and their dynamical laws have to be known (and monitored). It is this aspect of nonequilibrium response theory (the dynamical details mentioned in the abstract) which often restricts its applicability to systems with small number of degrees of freedom, and has prevented application to many body systems.

A general route for many body systems identifies a relevant subset of important (slow) degrees of freedom, and less relevant (fast) degrees are integrated out. Examples are the so called Mori-Zwanzig projection formalism \cite{Zwanzig,Zwanzig60,Zwanzig61,Mori58,Mori65} or Fokker-Planck- or Langevin equations \cite{risken, dhont}. When applying such approaches to nonequilibrium cases, the integrated degrees of freedom are typically assumed to be in equilibrium.

In this Letter, we derive a response scheme which overcomes these issues: Starting from the microscopic description, we derive a nonlinear response relation for a small subset of coarse grained degrees of freedom, which is then used in an extrapolation scheme: Measurements near equilibrium, i.e., linear in perturbation, are used to predict responses further away from equilibrium, i.e., to second order in perturbation. The microscopic degrees do neither have to be tracked, nor are they assumed to be equilibrating fast, so that this scheme is applicable to many body systems. We demonstrate applicability in an exactly solvable jump process and in computer simulations of the 2D Ising model.

{\bf Coarse grained nonlinear  response theory from path integrals --} We consider a classical many body system which is in weak contact with an equilibrium thermal bath. Considering for example the Ising model (see below), nonlinear response theory, as e.g. given in Ref.~\cite{Basu2015,PhysRevE.78.041120}, can only be applied if the Hamiltonian (e.g., nearest- or next to nearest neighbor interactions) and the dynamics (e.g., specific spin flip rules) are known, and if all degrees are tracked. Our goal is development of a nonlinear response method which can be applied by tracking a small number of degress of freedom, e.g., the order parameter in the Ising model, not necessitating knowledge about the details of the system.  

To this end, we introduce a coarse grained description in terms of $n$ (experimentally trackable) macrostates, each containing several, uniquely assigned microstates. At any time $t$, the system is thus characterized by a unique macrovariable  $X_t=0,1,\dots,n-1$ (e.g., the  sign of the magnetization in the Ising model is described by two macrostates, $X_t=0,1$) which represents the coarse grained phase space.  In the absence of perturbations, the system is in thermal equilibrium, and thus satisfies detailed balance and time reversal symmetry.

We aim to compute the response of the system to a perturbation, whose strength is quantified by the dimensionless parameter $\varepsilon$. The perturbation can for example be a force, an external field, or a change in the transition rates of a jump process. We restrict here to perturbations which are switched on at time $t=0$, but are otherwise time independent. We build on path integrals, in terms of which response theory has been worked out for the microsystem \cite{Colangeli11, Basu2015,Helden16}: The probability weight $p(\omega)$ of a microscopic path $\omega$ in the perturbed process differs from its equilibrium weight $p_{\rm eq}(\omega)$. This is captured by the action $\mathcal{a}(\omega)$, i.e., $p(\omega) = e^{-\mathcal{a}(\omega)}p_\text{eq}(\omega)$.  $\mathcal{a},$ which vanishes for $\ve=0$, is expanded in powers of $\ve$, 
\bea\label{eq:a}
\mathcal{a}=\varepsilon \left(\mathcal{d}'-\frac 12 \mathcal{s}'\right)+\frac{1}{2}\varepsilon^2 \mathcal{d}''+ \mathcal{O}(\ve^3) ,
\eea 
where the primes denote derivatives w.r.t. $\ve.$ In the spirit of Refs.~\cite{Baiesi09,Colangeli11,Basu2015},  $\mathcal{a}=\mathcal{d}-\frac{1}{2}\mathcal{s}$ is split into a part symmetric under time reversal, $\mathcal{d}$, and an antisymmetric part $\mathcal{s}.$ We take the perturbation to be such that $\mathcal{s}$ is linear in $\ve$, so that $\mathcal{s}''$ and higher derivatives vanish, which is a generic and useful case \cite{Kubo,Baiesi09,Seifert12}. This may also be interpreted as a definition of the order of perturbation: $\varepsilon$ is the quantity, in which $\mathcal{s}$ is linear. For a perturbation via potential forces this means that the perturbation Hamiltonian is linear in $\varepsilon$. 

The response of an observable, up to second order, can then be expressed in terms of equilibrium correlation functions involving  combinations of $\mathcal{s}'$ and $\mathcal{d}'$ \cite{Baiesi09,Colangeli11,Basu2015} (we will refer to the corresponding response formula when introducing Eq.~\eqref{eq:so} below).

On the {\it coarse grained} level the probability $P_{ij}$ of the macro-path which connects $X=i$ at $t=0$ and $X=j$ at time $t$ \footnote{The paths are sufficiently characterized by the parameters time $t$ and the initial and final states.}, follows from integration over  microstates, and the corresponding macro-action $\mathcal{A}_{ij}$ is (in the following, we omit the time arguments for brevity, keeping in mind that, e.g., $\mathcal{A}_{ij}=\mathcal{A}_{ij}(t)$) 
\begin{align}
\mathcal{A}_{ij} &\equiv  - \log \frac{P_{ij}}{P_{ij}^\text{eq}} =  \log\left[\frac{1}{P^{\rm eq}_{ij}}\int_{ij} d\omega~ p_\text{eq}(\omega)e^{-\mathcal{a}(\omega)}\right]. \label{eq:Aij}
\end{align}
Here,  $\int_{ij}$ denotes integration over only those micro paths $\omega$ which connect the macrostates $i$ (at $t=0$)  and $j$ (at time $t$). Using the definition, $\int_{ij} d\omega p_\text{eq}(\omega) =P_{ij}^\text{eq}$, the right hand side of Eq.~\eqref{eq:Aij} may be expanded in a series of $\ve$, to obtain the macroscopic analog of Eq.~\eqref{eq:a}. For $\ve=0$, the argument of the log is unity, and we use its expansion around that value to obtain,  with the notation $\mathcal{A}= \mathcal{D} - \frac{1}{2}\mathcal{S},$

\begin{align}
\mathcal{S}'_{ij} &\equiv \mathcal{A}'_{ji}-\mathcal{A}'_{ij}=\frac{1}{P^\text{eq}_{ij}}\int_{ij} \id\omega~ p_\text{eq}(\omega) \mathcal{s}'(\omega) ,\notag\\
\mathcal{D}'_{ij}& \equiv \frac{1}{2}\left(\mathcal{A}'_{ij}+\mathcal{A}'_{ji}\right)=\frac{1}{P^\text{eq}_{ij}}\int_{ij} \id\omega ~ p_\text{eq}(\omega) \mathcal{d}'(\omega),\notag\\
\mathcal{S}''_{ij}&\equiv\left(\mathcal{A}''_{ji}-\mathcal{A}''_{ij}\right) =2\mathcal{D}'_{ij}\mathcal{S}'_{ij}-\frac{2}{P^\text{eq}_{ij}}\int_{ij} \id\omega~ p_\text{eq}(\omega)  \mathcal{d}'\mathcal{s}'.\label{eq:spp}
\end{align}
The first derivatives, $\mathcal{S}'$ and  $\mathcal{D}'$, are thus given in terms of the microscopic counterparts, while notably, the coarse graining in general generates a finite $\mathcal{S}''$ in the last line of Eq.~\eqref{eq:spp}, although the microscopic counterpart $\mathcal{s}''$ is zero.

The expected value of a macro-observable $O(X)$ at time $t$ under the perturbation is given by the average over the macroscopic paths. Expanding $\mathcal{A}$ in powers of  $\ve$, we obtain, up to second order in $\ve$,
\bea
\la O(X_t) \ra&=& \sum_{ij}  P_{ij}O(j)=  \la O(X) \ra^{\rm eq} + \varepsilon \sum_{ij}\mathcal{S}'_{ij} P_{ij}^\text{eq} O(j) \cr
&-&\varepsilon^2 \sum_{ij} \mathcal{S}'_{ij} \mathcal{D}'_{ij} P_{ij}^\text{eq} O(j)+\frac{\varepsilon^2}{2}\sum_{ij}\mathcal{S}''_{ij} P_{ij}^\text{eq}O(j).\cr
&&\label{eq:soo}
\eea
Here $\langle\cdots \rangle$ and $\langle \cdots \rangle^{\rm eq}$ denote averages over the perturbed and equilibrium processes, respectively. Other terms in this expansion disappear because of time reversal symmetry of the equilibrium process,  manifest here in the symmetry of the matrix $P_{ij}^{\rm eq}$. The last term in Eq.~\eqref{eq:soo} is not present in the microscopic version \cite{Basu2015}, and it appears here because of the non-vanishing $\mathcal{S}''$ in Eq.~\eqref{eq:spp}.
The extrapolation scheme proposed below is applicable if the last term in Eq.~\eqref{eq:soo} vanishes. In particular, it is instructive to consider perturbations which couple to the coarse grained variable $X$. One example is a perturbation potential $\varepsilon V(X)$, i.e., a perturbation potential which is sensitive to the macrostates. In that case, $\mathcal{s}'(\omega)=\beta[V(X_0) - V(X_t)]$ \cite{Baiesi09},  with inverse thermal energy $\beta=(k_BT)^{-1}$. It is thus {\it equal} for all the micropaths connecting macro states $i$ and $j.$ 
Consequently, the term in the last line of Eq.~\eqref{eq:spp} simplifies to 
\bea
\int_{ij} \id\omega ~p_\text{eq}(\omega) \mathcal{d}'\mathcal{s}'= \mathcal{S}'_{ij}\int_{ij} \id\omega~ p_\text{eq}(\omega)  \mathcal{d}'=\mathcal{S}'_{ij}\mathcal{D}'_{ij}P^\text{eq}_{ij}.\;\quad \label{eq:sd}
\eea
It immediately follows that $\mathcal{S}''=0$ in Eq.~\eqref{eq:spp}, and therefore, Eq.~\eqref{eq:soo} simplifies to a form 
\bea 
\la O(X_t) \ra &=&\la O(X) \ra^{\rm eq} +\varepsilon \sum_{ij} \mathcal{S}'_{ij}  P_{ij}^\text{eq} O(j) \cr
&-& \varepsilon^2 \sum_{ij} \mathcal{S}'_{ij}  \mathcal{D}'_{ij} P_{ij}^\text{eq} O(j). \label{eq:so}
\eea
Equation~\eqref{eq:so}, an intermediate result, is the response formula for the coarse grained phase space $X$. 
It is reminiscent of the microscopic version \cite{Basu2015}, however here we obtained it for the croase grained variables. The left hand side is the nonequilibrium average of $O(X_t)$, while the right hand side is an explicit expression in terms of the time dependent matrices $\mathcal{\mathcal{S}}'$, $\mathcal{\mathcal{D}}'$ and $P^\text{eq}.$ Important for this work is the interpretation of Eq.~\eqref{eq:so}: It is worth appreciating that the second order response, given by the last term of Eq.~\eqref{eq:so}, involves  $\mathcal{S}'$ and $\mathcal{D}'$, which are the changes of these matrices to {\it linear} order in $\varepsilon$. This leads to the main result of the paper: Measuring the linear response of the system, i.e., measuring $\mathcal{\mathcal{S}}'$ and $\mathcal{\mathcal{D}}'$, is sufficient to predict the second order response from Eq.~\eqref{eq:so}.

This extrapolation scheme does neither rely on the knowledge or tracking of integrated degrees of freedom, nor are they assumed to equilibrate fast (in contrast to Zwanzig-Mori approaches), and is thus applicable to many body systems with the caveat that the linear response needs to be measured. We illustrate the scheme in two examples.

\begin{figure}[t]
 \centering
\includegraphics[width=4.7 cm]{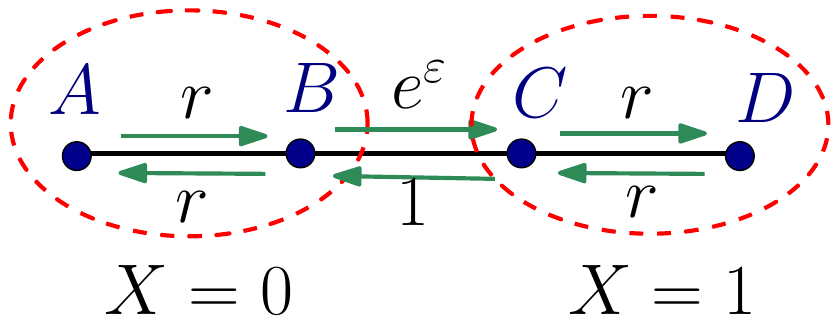}\\[3mm]%\hspace{.7cm}\includegraphics[width=3.7 
%cm]{./4st_cartoon2.pdf}\\ 
\includegraphics[width=8.8cm]{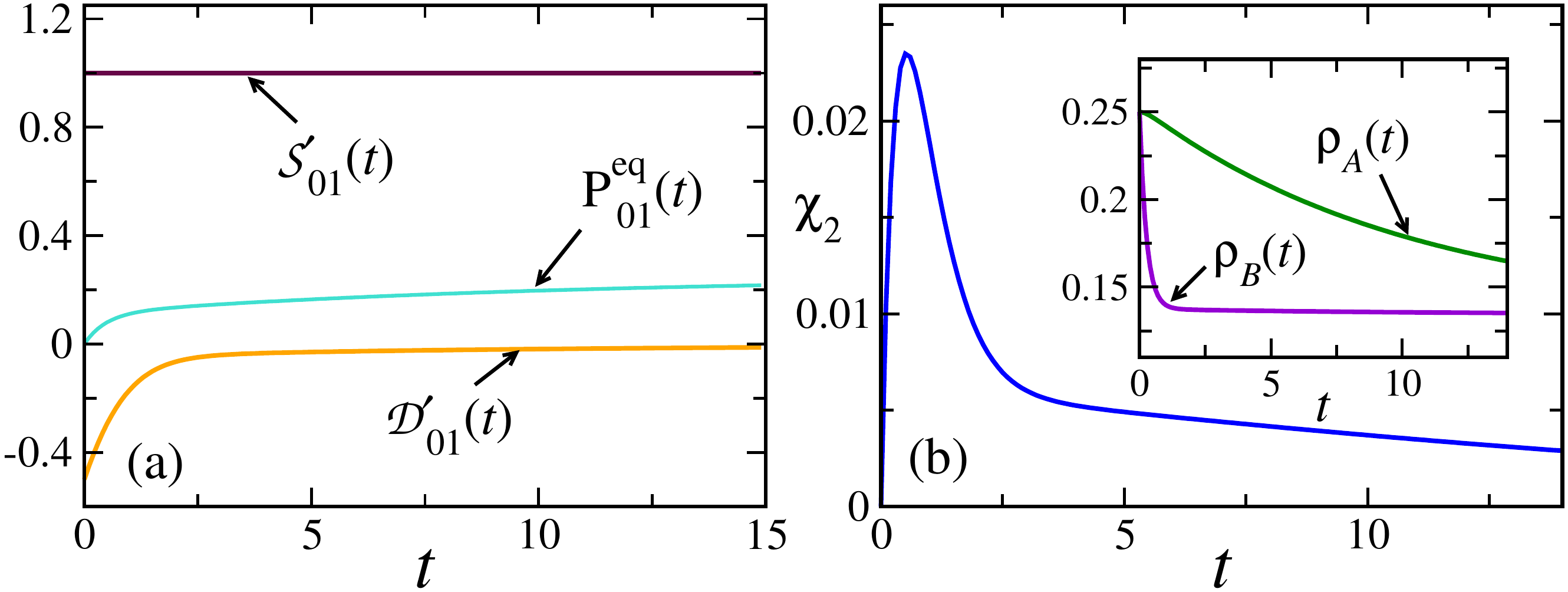} 
 % resp_c0_5.pdf: 360x252 pixel, 72dpi, 12.70x8.89 cm, bb=0 0 360 252
 \caption{Response in a coarse grained four state jump process as a function of dimensionless time $t$ after perturbing the center link,  for $r=0.1$. Microstates $A$ and $B$ are united to yield macro state $X=0$, $C$ and $D$ are merged to $X=1.$ (a) shows ${\cal S}'$, ${\cal D}'$ and $P^{\rm eq}$, the quantities of Eq.~\eqref{eq:chi_eq_2}.
(b) Second order response of $\la X\ra$ (see Eq.~\eqref{eq:def}).  
Inset gives the probabilities $\rho_{A/B}$ to find the system in state $A$ or $B,$ respectively, as a function of time.}
 \label{fig:4st_resp}
\end{figure}

{\bf  Four state jump process --} Let four  micro-states, $A,\dots,D$ be connected with given jump rates, see sketch in Fig.~\ref{fig:4st_resp}. The coarse grained macrostates combine $A,B$ ($X=0$) and $C,D$ ($X=1$), respectively, so that $X$ is the phase space of a two state system ($n=2$), with $\langle X\rangle^{\rm eq}=\frac{1}{2}$ because of symmetry.

At time $t=0$, the system is perturbed by switching the forward rate of the central link from 1 to $e^{\ve}$, while all other rates are left unchanged (see sketch in Fig.~\ref{fig:4st_resp}). Because we perturb the link connecting the macrostates, Eq.~\eqref{eq:so} can be used. We aim to find the responses up to the second order,
\begin{subequations}
\bea
\chi_1(t) &\equiv &\lim_{\ve\to0}\frac 1 {\ve} [\la O_t \ra - \la O \ra^\text{eq}],\label{eq:chi1}\\
\chi_2(t) &\equiv& \lim_{\ve\to0}\frac 1 {\ve^2}[\la O_t \ra - \ve \chi_1(t)- \la O \ra^\text{eq} ]. \label{eq:def}
\eea
\end{subequations}
The response formula, Eq.~\eqref{eq:so}, yields the predicted responses $\chi^\text{rf},$
\begin{subequations}\label{eq:8}
\begin{align}
\chi_1^{\rm rf} (t)&=  \sum_{ij} O(j)~ \mathcal{S}'_{ij} P_{ij}^\text{eq},\\ %= \mathcal{S}'_{01} P_{01}^\text{eq} ,\\ 
\chi_2^{\rm rf} (t)&=  -\sum_{ij} O(j)~ \mathcal{S}'_{ij}\mathcal{D}'_{ij} P_{ij}^\text{eq}. %=-\mathcal{S}'_{01} \mathcal{D}'_{01}  P_{01}^\text{eq} . 
\label{eq:chi_eq} 
\end{align}
\end{subequations}
Evaluating Eq.~\eqref{eq:8} in the extrapolation scheme, ${\cal S}'$ and  ${\cal D}'$ need to be known. Therefore, the path weight $P_{ij}(t)$ is measured in {\it linear} response (for $n=2$, a $2\times2$ matrix). Using Eqs.~\eqref{eq:Aij} and \eqref{eq:spp}, one then obtains, by employing also its equilibrium counterpart $P^\text{eq}_{ij}(t)$,  
\begin{subequations}\label{eq:A}
\bea
\mathcal{S}'_{ij} &=& \lim_{\ve\to0}\frac 1 {\ve} \log \frac{P_{ij}}{P_{ji}}, \\
\mathcal{D}'_{ij} &=& \lim_{\ve\to0}\frac 1 {2\ve}\log \frac{(P_{ij}^\text{eq})^2}{P_{ij} P_{ji}}.
\eea
\end{subequations}
The considered 4-state process is exactly solvable (see Supplemental Material \cite{Supp}), and $\mathcal{S}'$, $\mathcal{D}'$ and $P^\text{eq}$ so obtained are shown in Fig.~\ref{fig:4st_resp}(a). When applying the scheme experimentally, these curves are to be measured.

In this example, we take $O(X)=X$, i.e., we consider the response of $\langle X\rangle.$ The corresponding $\chi^\text{rf}$ are then found via Eq.~\eqref{eq:8} which, using $n=2$, simplifies to
\begin{subequations}
\begin{align}
\chi_1^{\rm rf} (t)&= \mathcal{S}'_{01} P_{01}^\text{eq} ,\\ 
\chi_2^{\rm rf} (t)&= -\mathcal{S}'_{01} \mathcal{D}'_{01}  P_{01}^\text{eq} . \label{eq:chi_eq_2}
\end{align}
\end{subequations}
Since $\mathcal{S}'_{ij}$ is anti-symmetric and we have $n=2$, the sums reduce to the term $0\to1$, and the nontrivial second order is the product of the functions shown in Fig.~\ref{fig:4st_resp}(a).  

We show analytically \cite{Supp} that Eq.~\eqref{eq:chi_eq_2} indeed yields the exact second order response, which, having coarse grained a four state to a two state model, is an explicit confirmation of the proposed scheme.

Fig.~\ref{fig:4st_resp} (b) shows the resulting $\chi_2$ as a function of time for an extreme choice of parameters: The rates $A\leftrightarrow B$ and $C\leftrightarrow D$ are {\it small} compared to the rates $B\leftrightarrow C$. Because of this, the density $\rho_A$ relaxes much slower than $\rho_B$ (inset), and the $\chi_2(t)$ shows two distinct time scales. This demostrates that Eq.~\eqref{eq:so} does neither rely on fast relaxation of integrated degrees, nor on Markovianity of the resulting two state system. For $t\to\infty$, $\chi_2$ vanishes because of symmetries.

{\bf 2d Ising model --} To demonstrate practical applicability, we consider Ising model on a  periodic square lattice with nearest-neighbor interactions among N spins $s_i=\pm 1$ and following Metropolis dynamics \cite{Newman1999}, studied using Monte-Carlo simulations \footnote{A randomly selected spin flips with a rate $\min \{ 1, e^{-\Delta H/T} \}$
where $\Delta H$ is the change in the Hamiltonian due to the proposed flip.}. See Ref.~\cite{PhysRevE.78.041120} for nonlinear response theory in the Ising model.
The Hamiltonian 
\bea
H = -  \sum_{\{ ij\}} s_i s_j - h \sum_{i=1}^N s_i+ \ve \Theta(t) \sum_{i=1}^{\cal N} s_i,\label{eq:I}
\eea
is asymmetric due to the presence of a magnetic field $h$ (included to allow for a finite $\chi_2$). $\ve$ gives the strength of perturbation which acts on $\cN \le N$ spins, and the unitstep function $\Theta(t)=0$ if $t<0$ and $\Theta(t)=1$ otherwise. 
With $k_B=1$, $h$ and temperature $T$ are dimensionless. For $h=0$, the 2d Ising model shows a paramagnet-ferromagnet-transition at temperature $T_c \simeq 2.269$ \cite{Baxter}. 
Our finite system with a lattice of size $N=16 \times 16$ and $T=2.45$ shows ferromagnetic order, however randomly flipping collectively the sign of the magnetization $m = \frac{1}{N}\sum_{i=1}^N s_i$, on a slow time scale. 

For the macrovariable $X=\sum_{i=1}^{\cal N} s_i$, corresponding to $n={\cal N}+1$ macrostates, the perturbation in Eq.~\eqref{eq:I} is of the form $V(X)$ (namely $V(X)=X$). An extreme limit is a {\it local} perturbation ($\cN=1$), where only a single tagged spin is perturbed. Here, the interpolation scheme is applied by only tracking (measuring) the dynamics of that tagged spin ($n=2$), while the configuration of the surrounding spins need not be known \footnote{We omit presentation of the numerical data for this case here.}.

\begin{figure}[t]
 \centering
 \centering{\includegraphics[width=6cm]{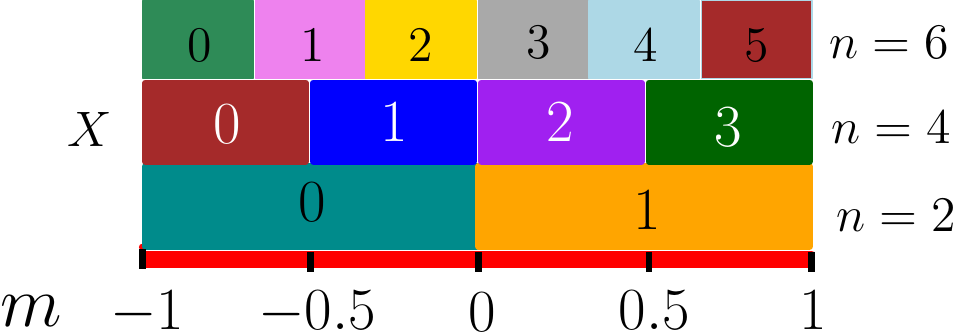}}

 \vspace{4mm}

 \includegraphics[width=8.8 cm]{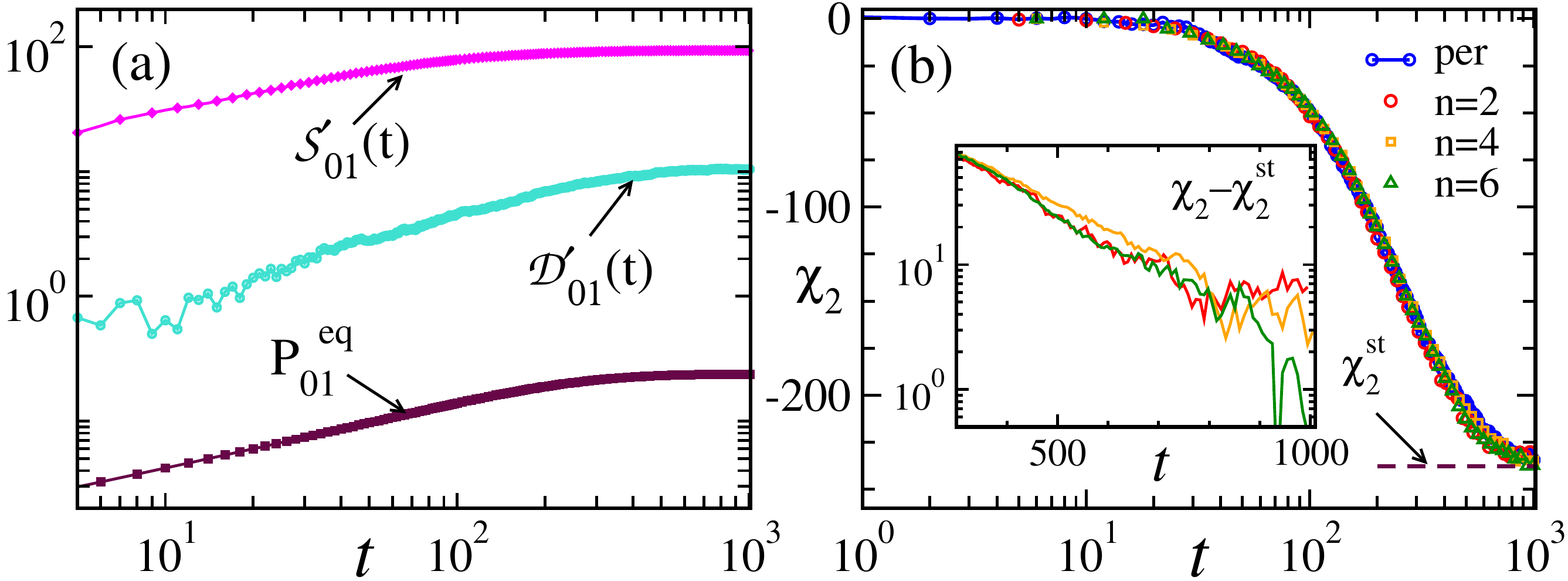}
\caption{Top: Sketch of the macrostates of the order parameter $m$ for different $n$. Bottom: (a) $\mathcal{S}_{01}'$ and $\mathcal{D}_{01}'$ (exemplarily shown for $n=2$),  measured at $\ve=0.0005$ along with  $P_{01}^\text{eq}$ as a function of time $t$ (in Monte-Carlo steps). (b) Second order response: Open symbols show $\chi_2^\text{rf}$, found using Eq.~\eqref{eq:chi_eq}, for the different values of $n$. The curve denoted `per' uses the conventional way [see Eq.~\eqref{eq:def}] of determining response functions for $\ve=0.003$. Horizontal dashed line gives the limit $\chi_2^\text{st}=\chi_2(t\to\infty)$ \cite{Supp}. Inset shows $\chi_2 - \chi_2^\text{st}$ (logarithmic scale). All curves are obtained from averaging more than $10^8$ trajectories. 
}
\label{fig:ising}
\end{figure}

More challenging, we consider a \emph{global} perturbation ($\cN=N$), aiming at the sign of the magnetization as  chosen observable of interest, specifically $O=\Theta(m)$. With $h=0.005$, $\langle O\rangle^{\rm eq} \simeq 0.613$ in the equilibrium state. Does one need $N+1=257$ macrostates in this case? Practically, a much smaller number turns out to be sufficient. We use $n=2$, $4$ and $6$ (see sketch in Fig.~\ref{fig:ising}), ruling out odd values for symmetry.

In our simulations, we measure ${\cal S}'$ and ${\cal D}'$ with a small value of $\varepsilon=0.0005$ using Eq.~\eqref{eq:A}~\footnote{To increase precision, all curves are measured at $\pm \ve$. The results for $\ve$ and $-\ve$  are added (linear in $\ve$) and subtracted (second order in $\ve$), making use of symmetry \cite{Supp,Helden16}.}. This yields the curves in Fig.~\ref{fig:ising}(a) (for ease of presentation, we only show the case $n=2$). The predicted second order response, $\chi^\text{rf}_2(t)$, is then given by Eq.~\eqref{eq:chi_eq}, i.e., summing over the matrix elements of ${\cal S}'$, ${\cal D}'$ and $P^{\rm eq}$. For $n=2$, this sum in given in Eq.~\eqref{eq:chi_eq_2}, and contains only one term: It is the product of the  functions in Fig.~\ref{fig:ising}(a). For larger $n$, more terms are summed. This yields the curves in Fig.~\ref{fig:ising}(b). We also measured the second order response using the conventional way (see Eq.~\eqref{eq:def}), for which we have used a larger value of $\ve=0.003;$  see the (blue) curve denoted `per' in Fig.~\ref{fig:ising}(b). The very good agreement in Fig.~\ref{fig:ising}(b) confirms the main claim of the paper: We used simulations at $\varepsilon=0$ and $\varepsilon=0.0005$, and obtained the nontrivial extrapolation to a larger perturbation  $\varepsilon=0.003$. As a practical aspect, the conventional way of determining $\chi_2$ (using Eq.~\eqref{eq:def}) needs about ten times the amount of computational effort to obtain curves with similar statistics. The curves for different $n$ can only be distinguished in a logarithmic presentation (Fig.~\ref{fig:ising}(b) inset), where the long time limit, found in a static measurement \cite{Supp}, is indeed aproached better and better for increasing $n$. We note that for other systems, the convergence with $n$ may be slower.

The scheme amounts to measuring transitions rates between the different values of $X$ which are, in suitable systems, obtained much more easily compared to the measuments needed for microscopic response theory.  Once experimental trajectories are obtained, the transition rates can be evaluated for different $n$, so that, larger $n$s do not necessarily require more experimental measuring time.

Recapitulating, $V=V(X)$ is a {\it sufficient} condition for accuracy of the proposed scheme. It means that {\it unperturbed} degrees of freedom can be coarse grained straightforwardly. In our examples, these are the unperturbed links or spins, but, in general, these can also include spatial or momentum degrees of freedom. Practically, we noted that the condition $V=V(X)$ is {\it not necessary}, so that much coarser descriptions as implied by this condition can suffice. By testing convergence with $n$, the accuracy of the method can be controlled. Because naturally, the obtained resolution for the observable is limited by the number of macrostates, this approach is especially useful if the behavior of a low dimensional observable is sought, such as the order parameter of a (phase) transition.

The presented coarse graining and extrapolation scheme constitute a conceptually new approach to nonlinear response theory. Because micro-degrees do not have to be monitored, it has a large range of applicability in complex systems. 
While circumventing the experimental need of applying strong perturbations, 
 the scheme can also be  more efficient regarding computation time compared to the traditional way of obtaining response functions, which is of additional advantage for slow systems. We note that at any order of perturbation, the response formula contains the change of $\mathcal{D}$ in at most one order lower, so that we expect the extrapolation to be  extendable beyond second order.

Future work will investigate time-dependent perturbations, and perturbations via nonconservative force fields. 

\begin{acknowledgements}
 We thank C. Maes for useful discussions. M.K. was supported by Deutsche Forschungsgemeinschaft (DFG)
Grant No. KR 3844/2-1.
\end{acknowledgements}

%One drawback of this method is that it is not easily applicable to time-dependent perturbations.} 

\bibliographystyle{apsrev4-1}
\bibliography{nLinResponse}

\newpage

\onecolumngrid
\begin{center}
 {\large \bf Supplemental Material for ``Extrapolation to nonequilibrium from coarse grained response theory'' }
\end{center}

%\vspace*{0.5 cm}

% \title{Supplemental Material for ``Extrapolation to nonequilibrium from coarse grained response theory''}
% 
% \author{Urna Basu}
% \affiliation{SISSA - International School for Advanced Studies and INFN,  Trieste, Italy}
% \author{Laurent Helden}
% \affiliation{2. Physikalisches Institut, Universit\"at  Stuttgart, 70550 Stuttgart, Germany}
% 
% \author{Matthias Kr\"uger}
% \affiliation{4th Institute for Theoretical Physics, Universit\"at
% Stuttgart, Germany} 
% \affiliation{Max Planck Institute for Intelligent Systems, 70569 Stuttgart, Germany}
%          
% 
% \maketitle
%\tableofcontents

%\section{Introduction}

%This document provides details of the exact solution of the four state jump process.

%\section{4-state jump process}

\twocolumngrid

{\bf 4-state jump process:} The 4-state jump process [refer to Fig. 1 in the main text], provides an example for which the nonlinear response can be exactly calculated, and exact validity of the extrapolation scheme can be demonstrated. 
The coarse grained path probabilities $P_{ij}(t)$ are expressed as sum over the microscopic paths connecting the macrostates $i$ and $j.$ For example, the path with initial state  $i=0$ and final state  $j=1$ at $t$ has a probability,
\bea
P_{01}(t) &=&\rho_A^\text{eq} [p_{AC}(t) + p_{AD}(t)] + \rho_B^\text{eq} [p_{BC}(t) + p_{BD}(t)] \cr
&& \label{eq:4st_P}
%P_{10}(t) =\rho_C (p_{CA}(t) + p_{CB}(t)) + \rho_D (p_{DA}(t) + p_{DB}(t)) \label{}
\eea
where $p_{\alpha\beta}(t)$  denotes the  probability that starting from micro-state $\alpha$ at $t=0$ the system reaches state $\beta$ in time $t$. We  obtained it exactly by solving the time dependent Master equation for any choice of jump rates; $\rho_\alpha^\text{eq}$ is the equilibrium probability for the system to be in the micro-state $\alpha.$

%As mentioned in the main text, the unperturbed system corresponds to the following choice of microscopic jump rates: $k_{AB}=k_{BA}=k_{CD}=k_{DC}=r.$ 

We consider, as in Fig. 1 in the main text, a perturbation which changes the `forward' jump rate connecting the two macrostates:  $k_{BC} =e^{\ve}.$ The matrices ${\cal S}'$ and ${\cal D}'$ are found form Eq. (8) in the main text, using $P_{ij}(t)$ from Eq.~\eqref{eq:4st_P}. Taking ratios of the equilibrium and perturbed macro-path probabilities, we obtain explicitly,
\bea
\mathcal{S}_{01}^\prime &=& 1 \cr
\mathcal{D}_{01}^\prime &=& \frac{e^{-2 \lambda t}[\lambda t (1+\lambda )(\lambda - r)-r] + \lambda t (\lambda -1)(\lambda +r) +r}{2 \lambda^2[e^{-2 \lambda t}(\lambda-r) - 2\lambda e^{(1+r-\lambda)t} +  \lambda + r]} \cr
&& \label{eq:4stSD1}
\eea
where $\lambda= \sqrt{1+r^2}.$ The second order derivative, $\mathcal{S}^{\prime \prime}=0$, vanishes exactly. %To calculate the 
The second order response of $\la X\ra,$ is then [following Eq. (9) in the main text] predicted,
\bea
\chi_2^\text{eq} = -\mathcal{S}_{01}^\prime \mathcal{D}_{01}^\prime P_{01}^\text{eq} \label{eq:chi2eq}
\eea
On the other hand, the second order response can also be found analytically without use of the response formula, by Taylor's expansion of the exact nonequilibrium result  $\la X(t) \ra = \rho_C(t) + \rho_D(t)$ where $\rho_\alpha(t)$ is the density in the micro-state $\alpha$ at time $t$ in the perturbed process. This yields
\bea
\chi_2^\text{}(t) &=& \frac {e^{-(1+r-\lambda)t}}{16 \lambda^3} \bigg [r(1-  e^{-2 \lambda t}) \cr
&+&  \lambda t \left \{ (\lambda -1)(\lambda +r)+(\lambda +1)(\lambda -r) e^{-2 \lambda t} \right\} \bigg ] \;\quad
\eea  
which matches exactly with Eq.~\eqref{eq:chi2eq} using  \eqref{eq:4stSD1}, so that $\chi_2^\text{eq}=\chi_2$ is demonstrated. \\

%\section{Static measurement of the second order response}

{\bf Static response formula for second order response:} In the case of a potential perturbation, the stationary long time values of the second order response $\chi_2$ can independently be obtained by a static response formula, which results from a Taylor's expansion of the Boltzman weight of the perturbed system.

%
% mesuring static correlation functions in the unperturbed state involving the perturbing potential and the observable. The relevant expression can be easily obtained using Taylor's expansion of the Boltzman weight of the perturbed system. 

For the Ising model discussed in the main text [see Eq. (11) therein], and for the case ${\cal N}=N$ (i.e., the global perturbation discussed in the main text),  second order response for observable $O$ is given by,
\bea
\lim_{t\to\infty}\chi_2 \equiv \chi_2^\text{st}= \beta^2 \left[\frac 12 \la m^2 ; O \ra - \la m\ra \la m; O\ra   \right], \label{eq:s}
\eea
where $m$ is the magnetization; $\la \cdot \ra$ denotes expectation in the unperturbed state and $\la A;B\ra = \la AB \ra -\la A \ra \la B\ra$ denotes the connected correlation.
%
% \begin{align}
% &\lim_{t\to\infty}\chi_2 =\notag\\&  \beta^2 \left[\frac 12 \big (\la m^2 O \ra -\la m^2 \ra \la O \ra \big) + \la m\ra^2 \la O\ra - \la m\ra\la mO \ra  \right]. \n %\label{eq:s}
% \end{align}
The above expression is used to compute $\chi_2^\text{st}$ which is displayed in Fig. 2 of the main text. We observe that this method yields a rather precise result which is used as a long time benchmark for the time dependent solutions.

{\bf Improving accuray of response data:}
Fig. 2 in the main text compares different methods (different numbers of macrostates) for finding second order responses. In order to obtain accurate data for this comparison, and to minimize sources of errors beyond the coarse graining, we measured all perturbed states under perturbations of $\ve$ as well as $-\ve$. We denote the corresponding transition rates $P_{ij}^{\ve}$ and $P_{ij}^{-\ve}$. Eq.~9 in the main text transforms then to
%o achieve better numerical accuracy removing possible error from higher (even) order effects we have extracted $\mathcal{S',D'}$ using measurements at two different values of perturbation, namely, $\pm \ve.$  This amounts to measuring $P_{ij}^{\ve}$ and $P_{ij}^{-\ve};$ using Eq. 9 in the main text, it is easily found that,   
% \bea
% \mathcal{S}'_{ij} &=& \frac 1 {2\ve} \left[ \log \frac{P_{ij}^\ve P_{ji}^{-\ve}}{P_{ij}^{-\ve} P_{ji}^{\ve}}  \right] \cr
% \mathcal{D}'_{ij} &=& \frac 1 {4\ve} \left[ \log \frac{P_{ij}^{-\ve} P_{ji}^{-\ve}}{P_{ij}^{\ve} P_{ji}^{\ve}}  \right]\label{eq:S6}
% \eea
\bea
\mathcal{S}'_{ij} =&\frac 1 {2\ve} \left[ \log \frac{P_{ij}^\ve P_{ji}^{-\ve}}{P_{ij}^{-\ve} P_{ji}^{\ve}}  \right]; 
\mathcal{D}'_{ij} = \frac 1 {4\ve} \left[ \log \frac{P_{ij}^{-\ve} P_{ji}^{-\ve}}{P_{ij}^{\ve} P_{ji}^{\ve}}  \right]\label{eq:S6}
\eea
These expressions have an error of order $\mathcal{O}(\ve^2)$, while Eq.~9 in the main text has an error $\mathcal{O}(\ve)$.  Eq.~\eqref{eq:S6} has been used to evaluate $\mathcal{S',D'}$ to find the curves in Fig. 2 in the main text.

Similarly, the directly measured second order response for $\la O\ra$ has been obtained from 
\bea
\chi_2^\text{per} = \frac 1{2\varepsilon^2} [\la O\ra_\varepsilon +  \la O\ra_{-\varepsilon} -2 \la O\ra_\text{eq}]\label{eq:S7}
\eea
where $\la O\ra_{\varepsilon}$ and $\la O\ra_{-\varepsilon}$ denote the expectation value of $O$ at perturbation strengths $\varepsilon$ and $-\varepsilon,$ respectively. Also for Eq.~\eqref{eq:S7}, the error ($\mathcal{O}(\ve^2)$) is reduced compared to Eq.~(7b) in the main text, which generaly yields an error of order $\mathcal{O}(\ve)$.

Using Eqs.~\eqref{eq:S6} and \eqref{eq:S7} thus improves accuray of the obtained curves in Fig. 2. We note that this method contains no principle change in strategy, and does not render the comparison of the proposed scheme to the conventional method, as it improves the data in both methods equally (changing errors from order $\mathcal{O}(\ve)$ to $\mathcal{O}(\ve^2)$). The formulas presented in the main text yield good data as well, however, the difference between different numbers of macrostates $n$ would be less easily apparent. Discussing and testing the behavior of the extrapolation scheme as a function of $n$ is an important aspect of this manuscript, so that we seek data as accurate as possible.

\end{document}